\documentclass{phb-proc4-auth}
\usepackage{graphicx}
\usepackage{amssymb}
\begin{document}
\begin{frontmatter}
\journal{SCES '04}
\title{Quasiparticle scattering in  superconductors}
\author[IP,CU]{Zden\v{e}k Jan\accent23 u\corauthref{1}\thanksref{ABC}}
\author[IP]{Franti\v{s}ek Soukup}
\author[IP]{Rudolf Tich\'{y}}
\author[IP,US]{Georgy Tsoi}
\author[TU]{Jan Hada\v{c}}
\author[CU]{Jana Vejpravov\'{a}}
\address[IP]{Institute of Physics AS CR, Na Slovance 2, 182 21 Prague, Czech Republic}
\address[CU]{Faculty of Mathematics and Physics, Charles University, Ke Karlovu 3, 121 16 Prague}
\address[US]{Department of Physics and Astronomy, Wayne State University, Detroit, MI 48202, USA}
\address[TU]{Faculty of Nuclear Sciences and Physical Engineering, Technical University of Prague, B\v{r}ehov\'{a} 7, 115 19 Prague}
 \thanks[ABC]{This work
was supported by the GACR 
(No. 102/02/0994) and by projects No. AVOZ1-010-914 and K1010104.}
\corauth[1]{Corresponding Author: 
Email: janu@fzu.cz}
\begin{abstract}
We compare results of high-resolution magnetic flux
(susceptibility) measurements in very weak magnetic field,
performed of three classes of superconductors. They reveal
astonishing details of the transition to the superconducting
state. Although Pb behaves also on this scale according to BCS
predictions, La is more reminiscent of HTS materials, which
exhibit anomalous features. We suggest that known structure
peculiarities are due to a strong electron-lattice instability and
lead to a resonance electron back scattering.
\end{abstract}
\end{frontmatter}
The interaction that causes conventional superconductivity
originates from a charge fluctuation with a frequency just below a
characteristic frequency of the ionic lattice that excites in the
lattice a resonant sympathetic vibration that overcompensates for
the electronic charge. As a result, part of the interaction
between two electrons in the medium is a spatially short-ranged,
temporally retarded attraction \cite{Ambegaokar69}. The drift
current resulting from electron scattering, particularly its
dependence on temperature $T$, the applied field $H$, and the
frequency of $ac$ field $\omega$, may be observed detecting the
flux generated by current induced in the sample by applied field.
An elastic electron scattering gives rise to imaginary electrical
conductivity while an inelastic one to real (Ohmic) conductivity.

The method used here allows to resolve variation of the
"wave-vector" as low as $10^{3}$ m$^{-1}$ and energy per electron
in the order of 1 feV, which is six orders of magnitude better
than give contact methods. In infinitesimally thin sample (wire
oriented parallel with field), the magnetization produced by the
shielding current is $\nabla\times
\mathbf{M}=\mathbf{j}=(ne/m)\left(\hbar \mathbf{K} - e\mathbf{A}
\right)=i\omega\sigma \mathbf{A}$. The $\mathbf{K}(T)=\sum
\mathbf{k} c_{k-K}^{*} c_{k}$ is the temperature dependent
wave-vector of the paramagnetic quasiparticle counter-flow
current, and $\sigma(T)$ is electrical conductivity. Above $T_c$
the paramagnetic drift current cancels the diamagnetic one,
$\mathbf{j}=-(ne^{2}/m)\mathbf{A}$, which is linear in response to
applied field. A common interpretation of temperature dependence
of conductivity (susceptibility) of superconductors is based on
the temperature dependent superfluid density $n_{s}(T)$. However,
it turns out that measurements in very low magnetic fields can be
hardly interpreted this way. A more plausible concept is based on
the temperature dependent quasi-particle scattering
\cite{Bardeen57}.

In conventional BCS superconductors the electron-electron pairing
mechanism is mediated by low angle electron-phonon forward
scattering. The superconducting-normal (SN) transition in type-I
superconductors measured by $ac$ susceptibility (inductively) in
weak magnetic fields is very sharp. (In 6N pure long thin Ga
single crystal it is 90$\%$ complete in a temperature interval of
2$\times$10$^{-6}$ K \cite{Gregory68}.) Since near below $T_{c}$
the gap opens linearly with temperature, $\Delta(T)\approx
\Delta'(T_{c})(T-T_{c})$. The temperature dependent paramagnetic
current follows the number of thermally excited quasi-particles,
i.e., $\tanh (\Delta(T)/2k_{B}T)$. The data recorded on Pb sphere
fit this dependence with the $\Delta'(T_{c}) \approx$ 1.3 eV/K,
see Fig. 1. The transition width is $\delta T \approx$ 1 mK, that gives the 
$\Delta(0)=\Delta'(T_{c})\delta T \approx$ 1.3 meV, the value
compatible with values $E_{g}(0)=2\Delta (0)= 2.73$ meV obtained
by spectroscopy or specific heat measurement. In the case of Pb
slab, the broadening of transition may be attributed to a
demagnetization factor, which gives rise to non-homogeneous
$\Delta'(H,T)$.

La is commonly regarded as a type I superconductor. But its
behaviour somewhat differs from the BCS model. La has two
structure phases, fcc and hcp, with superconducting critical
temperatures 6 K and 4.88 K respectively. An iso-structural phase
transition, attributed to electronic and lattice instability, has
an underlying dynamic mechanism of electronic origin. The
electronic system is strongly coupled to lattice. The Fermi
surface (FS) is modified at a general point at the zone boundary,
thus involving a rather large degeneracy of inequivalent saddle
points passing through $E_{F}$ \cite{Picktett80}. A nesting
feature of the FS is reduced at higher temperature \cite{Wang86}.
This second-order phase transition could be described by a
zone-boundary soft phonon mode, that may ultimately result in the
static distortion. The modulated phase is, in general, an
incommensurate one, which, may eventually lock into a commensurate
modulation due to the freezing of a specific phonon and result in
higher-order (anharmonic) terms in the strain energy
\cite{Castan03}. Comparing to Pb, the temperature dependence of
imaginary part of $ac$ conductivity of La shows: i) linear
segments; ii) low temperature part fitting Lorentzian,
$\mathrm{Im}\sigma(T) \propto 1/((T-T_{c})^{2}+\Gamma^{2})$; iii)
at decreasing $H_{ac}$ the dips rise below $T_{c}$. They increase
in amplitude, shift toward $T_{c}$, and are accompanied by "noise"
in real part. These features are extrinsic to the vortex matter,
to BCS or Ginzburg-landau models, but they can be related to
displacive structural transition and modulated phases.

\begin{figure}
    \centering
    \includegraphics[scale=0.45]{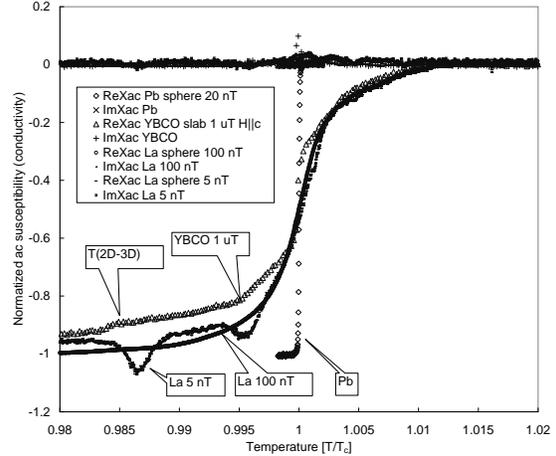}
    \caption{The SN transitions in Pb and La spheres and YBCO single-crystal slab oriented $c \parallel
\mathbf{H}_{ac}$. The transition in Pb falls in known data
\cite{Schooley72}. In YBCO transition occurs from 2D to 3D
superconductivity at temperature marked $T(2D-3D)$
\cite{Janu03A}.}

\end{figure}

In HTS cuprates, the temperature dependence of conductivity is
much more reminiscent of La than of Pb. The $\mathrm{Im}\sigma(T)$
may be well approximated by $\mathrm{Im}\sigma(T)\propto
\arctan(2(T-T_{c})/\Gamma)$, the Breit-Wigner resonance
\cite{Janu03B}. The $S$-shaped form with symmetry around $T_{c}$
and linear parts are inconsistent with the fluctuation model,
which is widely used to explain the convex part. With increasing
$H_{ac}$ the absorption peak on $\mathrm{Re}\sigma(T)$ broadens
and shifts toward the lower temperature whereas the opposite is
expected. The sharpness of the transition (10 mK at $T_{c}=91$ K)
suggests high underlying energies. We propose that this data may
be understood as resonance electron (back)scattering from
modulated lattice. Such a view is consistent with known data of
the FS properties and lattice structure available by other
experimental methods, which show strong electron localization,
stripes, CDW, SDW, pseudo-gap, real-space gap, and other
peculiarities \cite{Sato02,Kugler01,Zhu96,Kivelson03}.

We believe that because of the similar behavior of La and HTS, but
different from the character of Pb, the insight into the
diamagnetism of conduction electrons in "simple" (as compared with
HTS) La with its structure peculiarities, is relevant for
understanding of a mechanism of superconductivity in HTS. In
particular, we conjecture the relation between the correlated
electrons (pairing) and electron-lattice instability, and between
the elastic electron scattering and the soft mode or static
lattice distortion and strain.

Authors are grateful to L. Havela for stimulating discussion and
comments.

\end{document}